\documentclass[aps,onecolumn]{revtex4}
\usepackage{amsmath}
\usepackage{eurosym}
\usepackage{amssymb}
\usepackage{graphicx}
\usepackage{color}
\begin{document}
	
\title{Quark stars in generalized hybrid metric-Palatini gravity}
\author{Reyhaneh Aliannejadi}
\affiliation{School of Physics, Damghan University, Damghan, P.O.Box 36716-45667, Iran}
\author{Zahra Haghani}
\email{z.haghani@du.ac.ir}
\affiliation{School of Physics, Damghan University, Damghan, P.O.Box 36716-45667, Iran}

\begin{abstract}
 We investigate the physical properties of quark stars with two different equations of state in generalized hybrid metric-Palatini gravity. This theory corresponds to a bi-scalar gravitational theory in which there are two non-minimally-coupled scalar fields. The field equations of the metric tensor and scalar fields are derived by varying the action with respect to the dynamical fields.
We obtain the field equations for static spherically symmetric geometry interior of compact objects. The numerical solutions for two types of quark stars, the  MIT bag and CFL model are studied. The solutions reveal the structure of the compact objects. The solutions are obtained for different values of the model parameters. All considered cases show that compact objects are more massive than their general relativity counterpart. The surface redshift and compactness are also obtained which shows that the generalized hybrid metric-Palatini quark stars have higher surface redshift and are more compact than general relativity quark stars.
\end{abstract}

\maketitle

\tableofcontents

\section{Introduction}
The study of compact objects has become the interest of many theoretical physicists in recent years, especially after detecting the gravitational waves of merging binary systems of these objects \cite{GW}.  Investigating compact objects attracts attention from different aspects, and opens a new window to a field in which different branches of physics should have contributed to explain observational data. To obtain the equation of state of the compact objects, particle physics and quantum field theory is needed. Indeed at the center of compact objects such as neutron stars, the density is so high that one can consider the neutron star as a giant nuclei. So these objects can be considered as a natural laboratory of high energy physics. Also there are some motivations that maybe there are interactions between baryonic matter and dark matter at the core of compact objects. In ref. \cite{darkns1} ,\cite{darkns2} and \cite{darkns3} the possibility and results of such interaction are studied. 
On the other hand, the strong gravitational field of these objects can provide a new approach to test different gravity theories. The mass and radii obtained from the recent observational data do not match with the predictions of general relativity. The mass obtained from observational data has a wide range relative to the masses predicted by general relativity.  The Chandrasekhar mass limit for white dwarfs mass is about $1.4 M_\odot$ \cite{Chan}. On the other hand numerical and theoretical  investigations of the Tolman-Oppenheimer-Volkoff equation in ref  \cite{ruf} shows a limiting maximum mass of the neutron star of the order of $3.2 M_{\odot}$. This result obtained by using the stiff equation of state $p=\rho$. For a long time, observational data showed that the neutron stars should have the mass distribution centered on about $1.4 M_\odot$.

However,  the more accurate data obtained from astronomical observations and detecting of the gravitational waves  indicate that the mass range of neutron stars is wider from that obtained by applying the Chandrasekhar limit. The upper limit of the neutron star mass using gravitational and electromagnetic waves information  found in \cite{GW17} is $M_{max}<2.17M_\odot$.  Also, for the pulsar PSR J1614-2230, the obtained  mass is of order $1.928 ± 0.017M_\odot$ \cite{puls}. These and other observations of high mass neutron stars serve as motivations to explore new avenues for explaining this data. One possible way is to consider some exotic matters in the core of neutron stars. At very high densities,  theoretically suggested that the hadronic matter can undergo a phase transition to unpair quark matter \cite{wit}. Study of such matters in the core of neutron stars has captured significant attention \cite{quark1}. Such conditions can occur in two ways. The first is taking place in the quark-gluon phase transition in the early Universe. The second can be seen at ultrahigh densities in neutron stars where the neutron matter can convert into strange matter. One of the models to study the quark matter is the MIT bag model. In the bag model, we assume that quarks are confined to a sphere with certain radius (bag) and  in this region, quarks are considered as massless free particles \cite {MIT}. At ultrahigh densities, there is a possibility that the deconfiened quarks can form Cooper pairs near the Fermi surface. The pairing strength can be controlled by the gape parameter $\Delta$. This phenomena leads to the formation of Color-Flavor Locked (CFL) quark phase \cite{CFL}. The core of neutron stars is a suitable candidate to form the CFL quark phase. Hence, this model is very attractive from the astronomical and high-energy points of view. One can consider the neutron star as a quark matter and study the observation data to obtain some constraints on the parameters in the equation of the state of quark stars. There are many interesting works on the MIT and CFL quark stars in the  frame work of general relativity and modified gravity \cite{MICF}. In \cite{CFL-c}, it has been shown that the CFL stars  and their equation of state can successfully satisfy the constraints from gravitational wave observations.
With these new equations of state such MIT bag model, CFL model,  one can obtain higher maximum masses in the context of general relativity. 

Another way to explain observed  high  mass neutron stars is using modified gravity theories. This approach also has attracted many attentions recent years. There are a lot of works in which the mass of neutron stars are considered in different modified gravity theories \cite{modgrav}.  In this paper we will use a modified gravity theory known as generalized hybrid metric-Palatini gravity. In this model the gravitational action depends on a general function of both metric and Palatini curvature scalars  \cite{hybrid}. This theory is shown to be dynamically equivalent to the bi-scalar gravitational theory in which two scalar fields are non-minimally coupled. In ref. \cite{hybrid} the dynamical evolution of generalized hybrid-metric Palatini gravity is addressed. The authors studied in detail the late time acceleration of the universe for different potentials.  In the present work, we will consider the physical properties of compact objects in the context of the generalized hybrid metric-Palatini model. We use the MIT bag equation of state to obtain the compact object's interior solution. The physical conditions of the solution are considered and the results are compared with the results of general relativity.

This paper is organized as follow. In Sec. \ref{s1} the action of the theory and the field equations are obtained. In Sec. \ref{s2}, the field equations for a static and spherically symmetric compact objects are obtained. The solutions are obtained in Sec. \ref{s3} for the MIT bag  and CFL equations of state. In this section the mass-radius relation  for MIT bag and CFL quark stars are obtained for different values of the parameters. The results are compared with corresponding compact objects in  general relativity and observational mass data. Finally, in Sec. \ref{s4}, we conclude and discuss the results. 

\section{Action and Field Equations}\label{s1}
Let us consider the following action

\begin{equation}\label{act}
S=\int d^4x \sqrt{-g}f(R,\mathcal{R}),
\end{equation}
which is a generalization of the hybrid metric-Palatini gravity \cite{hybrid}. In the action \eqref{act}, the function 
$f(R,\mathcal{R})$ is an arbitrary function of the  Ricci scalar $R$ constructed by the Levi-Civita connection and  curvature scalar $\mathcal{R}$, formed by an independent connection denoted by $\hat{\Gamma}^{\alpha}_{\mu\nu}$. In \cite{gen}, the authors have shown that the action \eqref{act} is dynamically equivalent to the action with two scalar fields in the Einstein frame as
\begin{align}\label{fact}
	S=\int d^4 x \sqrt{-g} \Big(\frac{1}{2k^2} R -\frac{1}{2} (\nabla \phi)^2 -\frac{1}{2}e^\frac{-\sqrt{2} \kappa\phi}{\sqrt{3}} (\nabla \xi)^2 - W(\phi,\xi) \Big),
\end{align}
where $W(\phi,\xi)$ is an arbitrary interaction between two scalar fields $\phi$ and $\xi$. The action \eqref{fact} is known as Brans-Dicke or two-field inflation, the Brans-Dicke field is presented by the scalar field $\phi$ and the scalar field $\xi$ is dedicated for the inflaton. Our starting point in this work is the action \eqref{fact}. To investigate the physical structure of the compact objects we want to study the interior solution of the stars. So we need to add the matter action $S_m$  which is given by
\begin{align}
	S_m=\int d^4x \sqrt{-g}L_m,
\end{align}
to the action \eqref{fact}, where $L_m$ is the matter Lagrangian. The field equation can be obtained by varying the action with respect to the independent dynamical variables, i.e. metric tensor and the scalar fields,  $\phi$ and $\xi$.  The metric field equation can be obtained as
\begin{align}\label{eomm}
	G_{\mu\nu}=\kappa^2\Big(T_{\mu\nu}+T_{\mu\nu}^{(\phi)} +e^\frac{-\sqrt{2} \kappa\phi}{\sqrt{3}} 	T_{\mu\nu}^{(\xi)}- g_{\mu\nu} W \Big),
\end{align}
where $T_{\mu\nu}$ is the energy-momentum tensor of matter defined by
\begin{equation}
	T_{\mu\nu}=-\frac{2}{\sqrt{-g}}\frac{\partial\big(\sqrt{-g}L_m\big)}{\partial g^{\mu\nu}},
\end{equation}
and we have defined 
\begin{align}
	T_{\mu\nu}^{(\phi)}= \nabla_{\mu}\phi \nabla_{\nu}\phi - \frac{1}{2}  g_{\mu\nu} (\nabla\phi)^2,
\end{align}
and
\begin{align}
	T_{\mu\nu}^{(\xi)}=\nabla _{\mu}\xi \nabla_{\nu}\xi - \frac{1}{2}  g_{\mu\nu} (\nabla\xi)^2.
\end{align}
The field equations of the scalar fields are given by the following coupled equations
\begin{align}\label{eomfi}
	\Box \phi+\frac{\kappa}{\sqrt{6}} e^{-\kappa\phi \sqrt{2/3}}\left(\nabla \xi\right)^2-W_{\phi}=0,
\end{align}
\begin{align}\label{eomxi}
	\Box \xi+\frac{\kappa\sqrt{2}}{\sqrt{3}} \nabla_\mu \xi\,\nabla^\mu \phi-e^{\kappa\phi \sqrt{2/3}}W_{\xi}=0,
\end{align}
where we have used the notation $W_X=\frac{\partial W}{\partial X}$.

In the next section we will derive the field equations in the interior of  static and spherically symmetric objects. To solve these equation we will consider two different equations of state for the matter inside compact object.

\section{Field equations for static and spherically symmetric compact objects}\label{s2}
To study the properties of compact objects we will consider the static spherically symmetric line element given by
\begin{equation}
	ds^2=-A(r)dt^2+\frac{1}{1-2m(r)/r}dr^2+r^2d\Omega^2,
\end{equation}
where the metric components $g_{tt}$ and $g_{rr}$ are functions of the radial component $r$ that guarantee the static and spherical symmetry inside the compact object. The energy-momentum tensor of the matter is 
\begin{equation}
T_{\mu\nu}=(\rho+p)u_\mu u_\nu+p g_{\mu\nu},
\end{equation}
where $\rho=\rho(r)$ is the energy density and $p=p(r)$ is the pressure which both of them are only function of the radial coordinate due to the symmetries of the compact object. In the following, we have used the relativistic units, i.e. $c=1$ and $\kappa^2=8\pi G=1$. 

With the above assumptions for the metric  and energy-momentum tensors, one can obtain the components of the field equations as
\begin{align}
	\frac{2 m'}{r^2}=\rho +W+H_1,
\end{align}
\begin{align}
\frac{A'}{A r}\left(1-\frac{2 m}{ r}\right)-\frac{2
	m}{r^3}=p-W+H_1,
\end{align}
and
\begin{align}
		\left(1-\frac{2 m}{ r}\right)\left(\frac{A^{\prime\prime}}{2A}-\frac{A^{\prime2}}{4A^2}\right)-
	\frac{1}{r^2}\left(m^\prime -\frac{m}{r}\right)-\frac{A^\prime}{2rA }\left(m^\prime +\frac{m}{r}-1\right)=p
	-W-H_1,
\end{align}
where the symbol prime is denoted the derivative with respect to the radial coordinate ,$r$, and we  have also defined 
\begin{align}
	H_1=\frac12\left(e^{\sqrt{\frac{2}{3}} \Phi } \xi '^2+\Phi '^2\right)\left(1-\frac{2 m }{ r}\right).
\end{align}
It should be noted that, we have assumed that scalar fields,  $\phi$ and $\xi$, are functions of the radial coordinate.
In the above field equations, the contribution of the generalized hybrid metric-Palatini gravity is  summarized in the $W$ and $H_1$ terms. In the absence of these terms, the field equations reduce to the corresponging general relativity equations. The field equation for the scalar fields \eqref{eomfi} and \eqref{eomxi} in our case have the form
\begin{align}\label{emfi}
\left(1-\frac{2 m}{r}\right) \Phi ''+H_2\,	\Phi ' +\frac{e^{-\sqrt{\frac{2}{3}} \Phi }}{\sqrt{6}
		} (1-\frac{2m}{r}) \xi '^2-W_{\Phi }=0,
\end{align}
\begin{align}\label{emxi}
\left(1-\frac{2 m}{r}\right) \xi
''+H_2	\,\xi ' +\sqrt{\frac{2}{3}} (1-\frac{2 m}{r}) \xi ' \Phi '-e^{\sqrt{\frac{2}{3}} \Phi } W_{\xi }=0,
\end{align}
where
\begin{align}
	H_2=\frac{2}{r}-  \frac{m'}{r}-\frac{3 m}{r^2}  +\frac{A'}{2 A }(1-\frac{2m}{r})
\end{align}
The conservation equation of the energy-momentum tensor has only one non-zero component as 
\begin{align}
p^\prime+\frac{A'}{2 A }\left(\rho+p\right)&+H_2 \Phi'^2-\frac{2}{\sqrt{6}
} \cosh\left(\sqrt{2 /3}\Phi \right) (1-\frac{2m}{r}) \xi '^2 \Phi'\nonumber\\&+\left(e^{2 \sqrt{\frac{2}{3}} \Phi }-1\right) \xi ' W_{\xi }=0,
\end{align}
where we have used the scalar field equations \eqref{emfi} and \eqref{emxi}. In the next sections we will consider the case
\begin{align}\label{w}
	W\left(\Phi,\xi\right)=\alpha\left(\kappa \xi\right)^\lambda e^{-\lambda\kappa\Phi/\sqrt{6}},
\end{align}
where $\alpha$ and $\lambda$ are constant parameters. The cosmological evolution of the Universe in generalized hybrid metric-Palatini gravity for this choice of the interaction term \eqref{w} is studied in ref. \cite{hybrid}.

The obtained field equations are coupled and complicated. Therefore we will numerically  investigate them to determine the physical properties of the compact objects. To derive a quantitative  analysis of the field equations we will employ dimensionless parameters defined as
\begin{align}
p=\rho_c \,\bar{p},\quad \rho=\rho_c\,\bar{\rho},\quad \bar{m}=\sqrt{\rho_c}\,m, \quad \bar{r}=\sqrt{\rho_c}\,r,\quad\alpha=\rho_c\bar{\alpha}.
\end{align}
The scalar fields $\Phi$ and $\xi$ are dimensionless in relativistic units. In the next section, we will derive the physical  properties of two different compact objects by imposing the equation of state for them.
\section{Quark stars}\label{s3}
The physical properties of compact objects crucially depend on the equation of state. The core density of the neutron stars is higher than that of atomic nuclei, making the study of compact objects a natural laboratory for exploring such high densities. On the other hand, since we do not have access to such high densities at our laboratories, understanding the behavior of matter under these conditions is not practical. There is significant ongoing research to understand the equation of state of matter at high densities and temperatures.  In the following subsections, we will consider two different equations of state known as the MIT bag and the CFL equations of state. We will use numerical methods to solve field equations. The boundary condition  $m(r=0)=0$ implies that the mass of the star at its center is zero, which is logically expected. Additionally, $p(R)=0$ determines the radial coordinate of the star's surface. To solve the field equations, we will employ the shooting method. These equations will be numerically solved for a known central pressure $p(r=0)=p_c$ which is in the range of the predicted central pressure of neutron stars. Solving the equation $p(r=R)=0$ will yield the radial coordinate of the star's surface.

\subsection{MIT bag quark stars}
 One of the models for studying quark matter is the MIT bag model. In this model, it is assumed that quarks are confined to a sphere with a certain radius (the bag), and within this region, quarks are considered as massless free particles. The thermodynamics of  quark-gluon plasma is described by the equation of state
\begin{align}\label{mit}
	3 p=\rho-4 B,
\end{align}
where $B$ is the bag constant. To stabilize the system, there should be a vacuum pressure on the bag surface, one can interpret the bag constant $B$ as this pressure.
The equation of state \eqref{mit} can be obtained using finite temperature  quantum field theoretical methods \cite{MIT}. 	

In the following, we will consider quark stars described by the MIT bag equation of state \eqref{mit}  with  bag constant  $B=1.05\times10^{14}\,g/cm^3$ and central density $\rho_c=1.25\times 10^{15} \, g/cm^3$,  for four different values of the constant parameters in the frame work of the generalized hybrid metric-Palatini gravity.

\begin{figure}[htbp]	
	\centering
	\includegraphics[width=6.6cm]{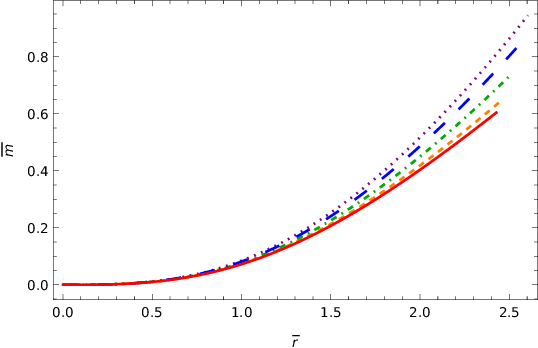}\hspace{.4cm}
		\includegraphics[width=6.7cm]{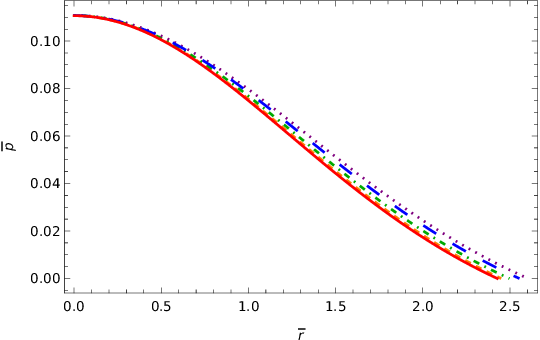}
	\caption{The interior mass profile (left panel) and interior pressure profile (right panel) of the MIT bag quark stars for four different values of the	constants $\bar{\alpha}$ and $\lambda$:  $\bar{\alpha}=0.01$  and  $\lambda=0.03$  (dashed curve),  $\bar{\alpha}=0.03$  and  $\lambda=0.03$ (dot-dashed curve), and $\bar{\alpha}=0.05$  and  $\lambda=0.01$ (long dashed curve), and $\bar{\alpha}=0.02$  and  $\lambda=0.07$ (dotted curve). The solid curve represents the standard general relativistic  ones for MIT  bag quark stars. }\label{maspres-mit}
\end{figure}

In Fig. \ref{maspres-mit}, variations of the mass and pressure inside of the compact object in terms of the dimensionless radial coordinate are plotted. The mass is a monotonically increasing function of radius, it is zero at the star's center and reaches to its maximum value at the surface of star. The solid curve in this figure, is devoted for general relativity. The mass profile shows that the generalized hybrid metric-Palatini gravity can explain more massive compact objects relative to the general relativity with the same equation of state and central density. The pressure is a monotonically decreasing function of radius. It is maximum at the center and is zero at the star's surface, $p(R)=0$ which determines the star's radius, $R$. It should be noted that we have used a physical equation of state and in the obtained solutions,  the pressure is positive everywhere, therefore, all relativistic energy conditions automatically will be satisfied. Also, one can see that using the equation of state \eqref{mit}, the sound speed $v_s=dp/d\rho=1/3$ is always smaller than the light speed, so the causality preserves inside the star.

\begin{figure}[htbp]	
	\centering
	\includegraphics[width=6.8cm]{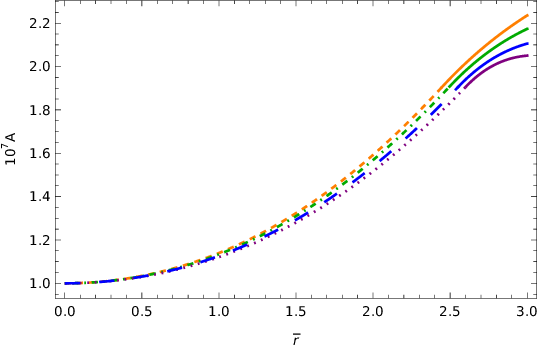}\\\vspace{.2cm}
	\caption{The variation of metric component $A$  in terms of the dimensionless radial coordinate of the MIT bag quark stars for four different values of the constants $\bar{\alpha}$ and $\lambda$:  $\bar{\alpha}=0.01$  and  $\lambda=0.03$  (dashed curve),  $\bar{\alpha}=0.03$  and  $\lambda=0.03$ (dot-dashed curve), and $\bar{\alpha}=0.05$  and  $\lambda=0.01$ (long dashed curve), and $\bar{\alpha}=0.02$  and  $\lambda=0.07$ (dotted curve). }\label{A-mit}
\end{figure}

In Fig. \ref{A-mit}, the variation of metric component $A$ in terms of the dimensionless radial coordinate $\bar{r}$  is plotted. The qualitative behavior of this function for different values of the constant parameters is the same. For each curve, the starting point for the solid line is the star's radius. Our results demonstrate the continuity of metric component at the stellar surface. This ensures a smooth matching of the interior and exterior solutions at this boundary. The initial conditions for exterior solutions are obtained from the interior solutions evaluated at $r=R$.

\begin{figure}[htbp]	
	\centering
	\includegraphics[width=6.8cm]{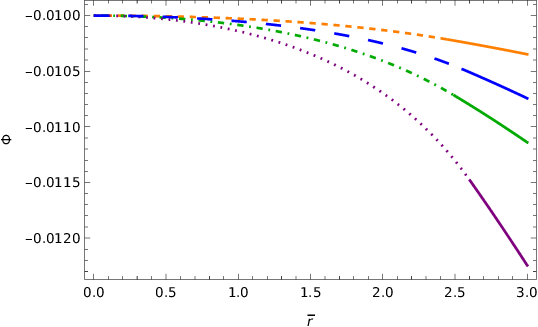}\hspace{.4cm}
	\includegraphics[width=6.6cm]{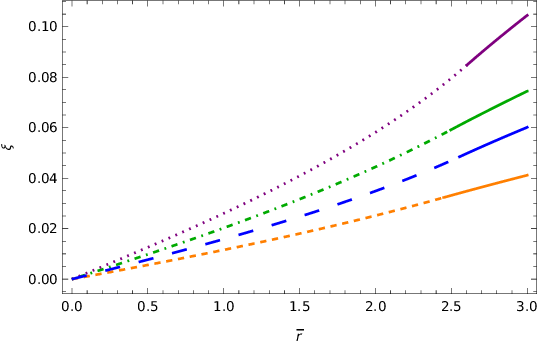}
	\caption{The variation of  scalar fields $\Phi$  (left panel) and $\xi$ (right panel) in terms of the dimensionless radial coordinate of the MIT bag quark stars for four different values of the	constants $\bar{\alpha}$ and $\lambda$:  $\bar{\alpha}=0.01$  and  $\lambda=0.03$  (dashed curve),  $\bar{\alpha}=0.03$  and  $\lambda=0.03$ (dot-dashed curve), and $\bar{\alpha}=0.05$  and  $\lambda=0.01$ (long dashed curve), and $\bar{\alpha}=0.02$  and  $\lambda=0.07$ (dotted curve).  }\label{phixi-mit}
\end{figure}

Fig. \ref{phixi-mit}  shows the variation of  scalar fields $\Phi$ and $\xi$, inside and outside the star. The behavior of scalar fields are smooth and well-defined throughout, demonstrating continuity at the surface of the star.  The solid portion of the graph depicts  the exterior behavior of the scalar field.

\begin{figure}[htbp]	
	\centering
	\includegraphics[width=6.6cm]{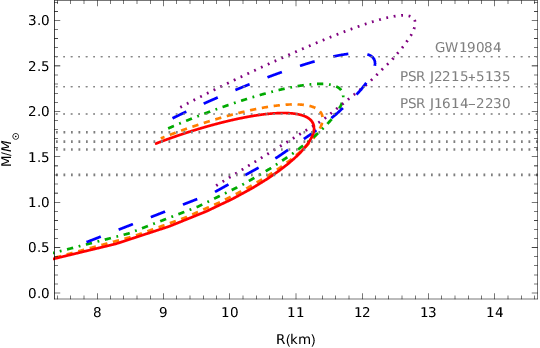}\hspace{.4cm}
		\includegraphics[width=6.6cm]{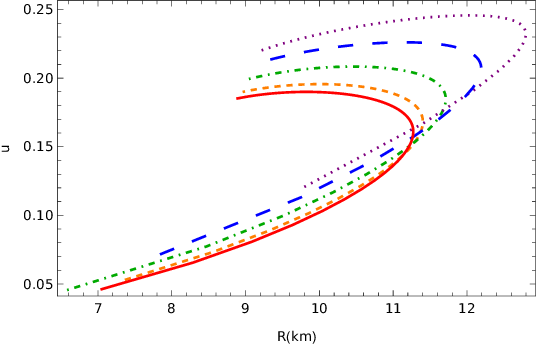}\hspace{.4cm}
	\caption{The mass-radius relation (left panel) and compactness (right panel) of the MIT bag quark stars for four different values of the constants $\bar{\alpha}$ and $\lambda$:  $\bar{\alpha}=0.01$  and  $\lambda=0.03$  (dashed curve),  $\bar{\alpha}=0.03$  and  $\lambda=0.03$ (dot-dashed curve), and $\bar{\alpha}=0.05$  and  $\lambda=0.01$ (long dashed curve), and $\bar{\alpha}=0.02$  and  $\lambda=0.07$ (dotted curve). The solid curve represents the standard general relativistic  ones for MIT  bag quark stars. The dotted horizontal lines show the mass of compact objects from observational data.}\label{mr-mit}
\end{figure}

The left panel of Figure~\ref{mr-mit} presents the mass-radius relation for MIT bag quark stars within the framework of generalized hybrid metric-Palatini gravity compared to general relativity (solid line).  The central density range considered here is $4.75\times10^{14}\,g/cm^3<\rho_c<1.05\times 10^{16}\, g/cm^3$. This figure clearly shows that the generalized hybrid metric-Palatini gravity can include a wider range of masses compare to general relativity. Therefore this model can predict more massive compact objects. The dotted horizontal lines in this figure represent the mass of stars from observational data. The mass obtained from GW19084 and PSR J2215+5135 are not in the range of masses predicted by MIT bag quark stars in general relativity, however, the presented model predicts the existence of even more massive objects than those currently observed.

 The compactness of a compact object is defined as the mass-to-radius ratio (denoted by $u$). Compactness is a crucial parameter that determines the strength of a compact object's surface gravitational field. For larger compactness values, we would expect more relativistic effects which can affect the trajectories of photons emitted from the compact object. By analyzing compactness, we can assess the validity of various gravitational theories in regimes of high gravity and potentially constrain the equation of state parameters for compact objects.  In the right panel of  Fig. \ref{mr-mit}, the compactness of the MIT bag quark stars in terms of their radius are plotted. The solid curve represents stars within the framework of general relativity, while the other curves depict stars under generalized hybrid metric-Palatini gravity with varying parameters ($\bar{\alpha}$ and $\lambda$). This figure shows that the MIT stars in the present model are more compact in comparison to the general relativity counterparts. Also, the compactness essentially depends on the value of model parameters.
\begin{table}[h!]
	\begin{center}
	\begin{tabular}{|c|c|c|c|c|}
		\hline
		$\lambda$&~~~$0.03$~~~&~~~$0.03$~~~&~~~$0.01$~~~&~~~$0.02$~~~ \\
		\hline
		$\bar{\alpha}$ &~~~$0.01$~~~&$~~~0.03~~~$&$~~~0.05~~~$&~~~$0.07$~~~ \\
		\hline
		\quad$M_{max}/M_{\odot}$\quad& $~~~2.07~~~$& $~~~2.3~~~$& $~~~2.75~~~$&~~~$3.24$~~~\\
		\hline
		$~~~R\,({\rm km})~~~$& $~~~10.98~~~$& $~~~11.37~~~$& $~~~12.29~~~$&~~~$13.14$~~~\\
			\hline
		$~~~~Z_s~~~~$& $~~~0.50~~~$& $~~~0.58~~~$& $~~~0.72~~~$&~~~$0.92$~~~\\
		\hline
		$~~~\rho_{c} \times 10^{-15}\,({\rm g/cm}^3)~~~$& $~~~1.84~~~$& $~~~1.5~~~$& $~~~1.06~~~$&~~~$0.74$~~~\\
		\hline
	\end{tabular}
	\caption{The maximum masses, surface redshift, the corresponding radii  and central densities for the MIT bag quark stars in generalized hybrid metric-Palatini gravity.}\label{mit-tab}
\end{center}
\end{table}

Table~\ref{mit-tab} presents the maximum masses and corresponding radii for selected values of the constant parameters.The value of  maximum mass essentially depends on the parameters $\alpha$ and $\lambda$. The central density for the maximum mass is of order $10^{15}g/cm^3$.  The table also includes the surface redshift ($	Z_s$), defined as
\begin{align}
	Z_s=\left(1-\frac{2GM}{c^2 R}\right)^{-1/2}-1.
\end{align}
Surface redshift is one of the possible ways to distinguish different stellar models. Also with the help of this parameter maybe can put some constraint on the equation of state at high densities.

 For a more accurate comparison of the result of generalized hybrid metric-Palatini with general relativity, the maximum mass, radius, central density and surface redshift of the MIT bag star in the general relativity framework are  $M=1.98M_\odot$, $R=10.81\,km$,  $\rho_c=2.01\times10^{15}g/cm^3$ and $Z_s=0.48$.  
\subsection{CFL quark stars}
In this section we want to consider the CFL quark stars. The equation of state of the paired quark matter is given by
\begin{align}
	p=\frac{1}{3}\left(\rho-4 B\right)+\frac{3\alpha\delta^2}{\pi^2},
\end{align}
where 
\begin{align}
	\delta^2=-\alpha+\sqrt{\alpha^2+\frac{4}{9}\pi^2(\rho-B)},
\end{align}
and $\alpha=-\frac{m_s^2}{6}+\frac{2\Delta^2}{3}$.
In the case of unpaired ($\Delta=0$) and massless  ($m_s=0$) quarks, the CFL equation of state reduces to the MIT bag equation of state.
We investigate cases with different values of the model parameters $\alpha$ and $\lambda$  but one specific set of values of parameters of the equation of state in which $\Delta=250\, MeV$, $m_s=150\, MeV$ and $B=2\times 10^{14}g/cm^3$.
\begin{figure}[htbp]	
	\centering
	\includegraphics[width=6.5cm]{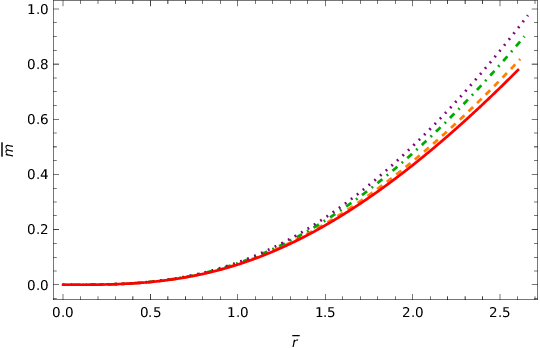}\hspace{.3cm}
	\includegraphics[width=6.6cm]{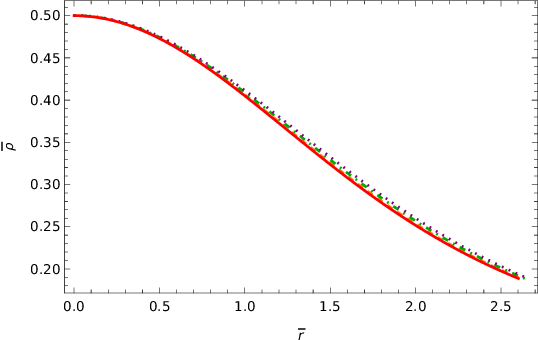}
	\caption{The interior mass profile (left panel) and interior density profile (right panel) of the CFL quark stars for three different values of the	constants $\bar{\alpha}$ and $\lambda$:  $\bar{\alpha}=0.01$  and  $\lambda=0.03$  (dashed curve),  $\bar{\alpha}=0.03$  and  $\lambda=0.03$ (dot-dashed curve), and $\bar{\alpha}=0.05$  and  $\lambda=0.05$  (dotted curve). The solid curve represents the standard general relativistic  ones for CFL quark stars. }\label{maspres-cfl}
\end{figure}

Fig. \ref{maspres-cfl} shows the variation of mass (left panel) and pressure (right panel) from center to the surface of the CFL quark star with central density $\rho_c=1.25\times10^{15}\, g/cm^3$. The different  curves are for different values  of the parameters $\bar{\alpha}$ and $\lambda$. The mass profile is a monotonically increasing function of the radial coordinate which reaches to its maximum value at the surface of the star. The pressure is maximum at the center and decreases monotonically until  zero. As one can see, the variation of mass essentially depends on the values of the parameters. All curves in the right panel show that the predicted mass for the CFL quark star in the generalized hybrid metric-Palatini gravity is greater than the corresponding case in general relativity.
\begin{figure}[htbp]	
	\centering
	\includegraphics[width=6.8cm]{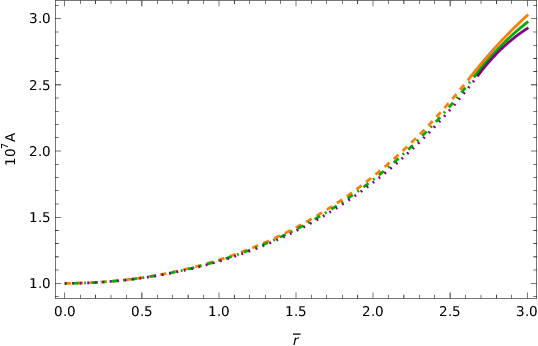}\\\vspace{.2cm}
	\caption{The variation of the metric component $A$  in terms of the dimensionless radial coordinate of the CFL quark stars for three different values of the	constants $\bar{\alpha}$ and $\lambda$:  $\bar{\alpha}=0.01$  and  $\lambda=0.03$  (dashed curve),  $\bar{\alpha}=0.03$  and  $\lambda=0.03$ (dot-dashed curve), and $\bar{\alpha}=0.05$  and    $\lambda=0.05$  (dotted curve). }\label{A-cfl}
\end{figure}
In Fig. \ref{A-cfl}, the variation of the $(00)$ component of the metric tensor $(A)$ in terms of the dimensionless radial coordinate for different values of the parameters is shown. As the radial coordinate increases, the value of $A$ augments. The solid part of the curves shows the exterior solution (where $\rho=0=p$) of the field equations. One can see that this quantity is continuous at the surface of the star. Near the center of the star, all curves almost coincide, regardless of the parameter values.
\begin{figure}[htbp]	
	\centering
	\includegraphics[width=6.8cm]{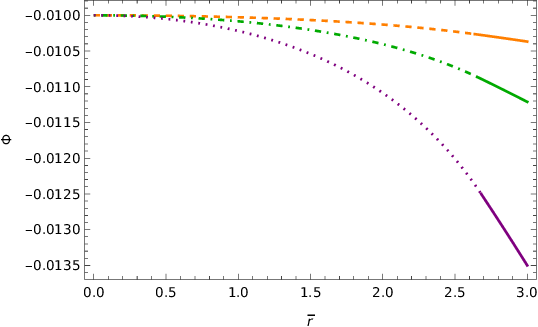}\hspace{.4cm}
	\includegraphics[width=6.5cm]{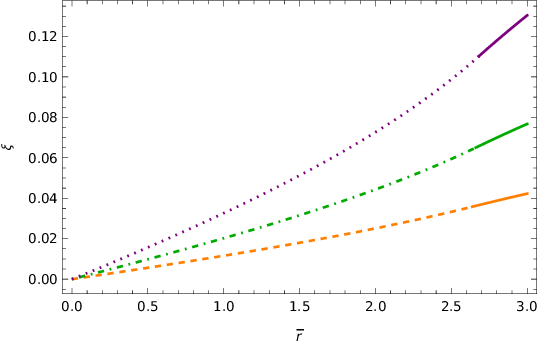}
	\caption{The variation of  scalar fields $\Phi$  (left panel) and $\xi$ (right panel) in terms of the dimensionless radial coordinate of the CFL quark stars for three different values of the	constants $\bar{\alpha}$ and $\lambda$:  $\bar{\alpha}=0.01$  and  $\lambda=0.03$  (dashed curve),  $\bar{\alpha}=0.03$  and  $\lambda=0.03$ (dot-dashed curve), and $\bar{\alpha}=0.05$  and   $\lambda=0.05$  (dotted curve). }\label{phixi-cfl}
\end{figure}
In Fig. \ref{phixi-cfl}, the behavior of the scalar fields $\Phi$ (right panel) and $\xi$ (left panel) inside and outside of the CFL star are shown. Each plot contains three curves corresponding to three different values of the parameters $\bar{\alpha}$ and $\lambda$. The variation of scalar fields crucially depends on the value of parameters. These scalar fields are continuous and smooth everywhere inside and outside of the star.  The scalar field $\Phi$ is a decreasing function of $\bar{r}$ while $\xi$ is an increasing function of the radial coordinate.`
\begin{figure}[htbp]	
	\centering
	\includegraphics[width=6.5cm]{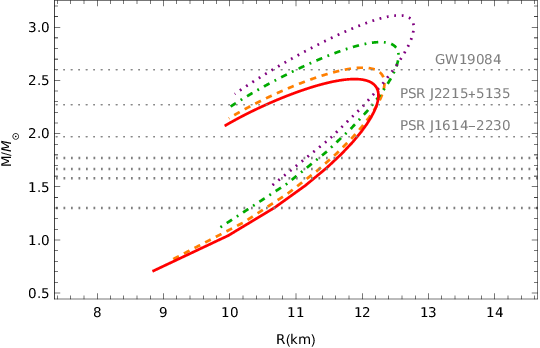}\hspace{.38cm}
	\includegraphics[width=6.5cm]{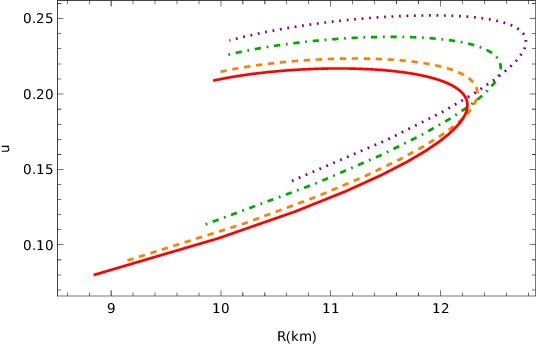}
	\caption{The mass-radius relation (left panel) and compactness  (right panel) of the CFL quark stars for three different values of the constants $\bar{\alpha}$ and $\lambda$:  $\bar{\alpha}=0.01$  and  $\lambda=0.03$  (dashed curve),  $\bar{\alpha}=0.03$  and  $\lambda=0.03$ (dot-dashed curve), and $\bar{\alpha}=0.05$    $\lambda=0.05$  (dotted curve). The solid curve represents the standard general relativistic  ones for CFL quark stars. The dotted horizontal lines show the mass of compact objects from observational data.}\label{mr-fl}
\end{figure}

In Fig. \ref{mr-fl}, the mass-radius (left panel) and compactness (right panel) of the CFL quark star are shown. The right panel shows that the range of the predicted mass of CFL quark stars in the framework of the generalized hybrid metric-Palatini is wider than the ones in GR. Also, the maximum mass crucially depends on the value of the model parameters. The left panel shows the compactness of the CFL stars. The CFL stars in the present model are more compact relative to the GR ones.
\begin{table}[h!]
	\begin{center}
		\begin{tabular}{|c|c|c|c|}
			\hline
			$\lambda$&~~~$0.03$~~~&~~~$0.03$~~~&~~~$0.05$~~~\\
			\hline
			$\bar{\alpha}$ &~~~$0.01$~~~&$~~~0.03~~~$&$~~~0.05~~~$\\
			\hline
			\quad$M_{max}/M_{\odot}$\quad& $~~~2.61~~~$& $~~~2.86~~~$& $~~~3.11~~~$\\
			\hline
			$~~~R\,({\rm km})~~~$& $~~~12.01~~~$& $~~~12.27~~~$& $~~~12.54~~~$\\
			\hline
			$~~~~Z_s~~~~$& $~~~0.68~~~$& $~~~0.79~~~$&~~~$0.93$~~~\\
			\hline
			$~~~\rho_{c} \times 10^{-15}\,({\rm g/cm}^3)~~~$& $~~~1.54~~~$& $~~~1.31~~~$& $~~~1.12~~~$\\
			\hline
		\end{tabular}
		\caption{The maximum masses, surface redshift and the corresponding radii  and central densities for the CFL quark stars in generalized hybrid metric-Palatini gravity.}\label{cfl-tab}
	\end{center}
\end{table}  

In Table \ref{cfl-tab},  the maximum masses and other properties of the CFL quark stars for different values of the parameters  $\bar{\alpha}$ and $\lambda$ are presented. To more accurate comparison of the result of generalized hybrid metric-Palatini with general relativity, the maximum mass, radius, central density and surface redshift of the CFL star in the framework of general relativity are  $M=2.51M_\odot$, $R=11.89\,km$,  $\rho_c=1.65\times10^{15}g/cm^3$ and $Z_s=0.63$. The surface redshift of the CFL stars depends on the value of the model parameters.

\section{Discussion}\label{s4}
In this paper, we have studied static and spherically symmetric compact objects in the context of a modified gravity theory known as the generalized hybrid metric-Palatini gravity. It has been shown that this theory is dynamically equivalent to a double scalar-tensor theory. The field equations are derived from the variational principle. In this theory, a term describes the interaction between two scalar fields. To solve the field equations one needs to know the interaction term and also the equation of state of matter content of the star. In the core of neutron stars, the density is so high that a hadron-quark transition can occur. We consider two different equations of state for the quark stars. In one of them, the quarks are considered as free and massless particles in a bag which is known as the MIT bag model.  However, at ultra-high densities, it is shown that quarks can form the Cooper pairs. The equation of state for this case is known as the CFL quark phase. In both cases, the physical properties of the quark stars are studied for different values of the parameters of the generalized hybrid metric-Palatini gravity and the results are compared to the general relativity ones. The stars in the present model are more massive and compact relative to the general relativity. Also, the surface redshift is higher than the ones in GR.



\end{document}